\documentclass{emulateapj}
\usepackage{apjfonts}
\usepackage{epsfig}

\renewcommand\u{{\vec{\bf u}}}
\newcommand\pg{\phi_\gamma}
\newcommand\dt{{\Delta\theta}}
\newcommand\that{\Delta\hat\theta}
\newcommand\BA{B{\hat A}}
\newcommand\avg[1]{\langle{#1}\rangle}
\newcommand\Lcut{L_{\rm cut}}

\newcommand\refeq[1]{eq.~(\ref{eq:#1})}
\newcommand\refsec[1]{\S~\ref{sec:#1}}
\newcommand\reffig[1]{Figure~\ref{fig:#1}}

\begin{document}

\title{Effects of Ellipticity and Shear on Gravitational Lens 
Statistics}

\author{
Dragan Huterer,\altaffilmark{1,2,3}
Charles R.\ Keeton\altaffilmark{3,4,5}
\& Chung-Pei Ma\altaffilmark{6}
}

\altaffiltext{1}{Department of Physics, Case Western
Reserve University, Cleveland, OH~44106}

\altaffiltext{2}{Kavli Institute for Cosmological Physics,
University of Chicago, Chicago, IL~60637}

\altaffiltext{3}{Astronomy \& Astrophysics Department,
University of Chicago, Chicago, IL~60637}

\altaffiltext{4}{Hubble Fellow}

\altaffiltext{5}{Physics \& Astronomy Department, Rutgers 
University, Piscataway, NJ 08854}

\altaffiltext{6}{Department of Astronomy, University of 
California, Berkeley, CA 94720}
 
\begin{abstract}
We study the effects of ellipticity in lens galaxies and external 
tidal shear from neighboring objects on the statistics of strong 
gravitational lenses.  For isothermal lens galaxies normalized so 
that the Einstein radius is independent of ellipticity and shear, 
ellipticity {\it reduces\/} the lensing cross section slightly, 
and shear leaves it unchanged.  Ellipticity and shear can 
significantly enhance the magnification bias, but only if the 
luminosity function of background sources is steep.  Realistic 
distributions of ellipticity and shear {\it lower\/} the total 
optical depth by a few percent for most source luminosity 
functions, and increase the optical depth only for steep 
luminosity functions.  The boost in the optical depth is 
noticeable ($\gtrsim$5\%) only for surveys limited to the 
brightest quasars ($L/L_* \gtrsim 10$).  Ellipticity and shear 
broaden the distribution of lens image separations but do not 
affect the mean.  Ellipticity and shear naturally increase the 
abundance of quadruple lenses relative to double lenses, 
especially for steep source luminosity functions, but the effect 
is not enough (by itself) to explain the observed
quadruple-to-double ratio.  With such small changes to the optical 
depth and image separation distribution, ellipticity and shear 
have a small effect on cosmological constraints from lens 
statistics: neglecting the two leads to biases of just 
$\Delta\Omega_M = 0.00\pm0.01$ and
$\Delta\Omega_\Lambda = -0.02\pm0.01$ (where the errorbars 
represent statistical uncertainties in our calculations).
\end{abstract}

\keywords{cosmology: theory --- gravitational lensing}

\section{Introduction}
\label{sec:intro}

A circularly symmetric gravitational lens is a useful theoretical 
construct that, most likely, will never be observed in a 
cosmological setting.\footnote{The only exception is Galactic 
microlensing, where the separation between stars is large enough 
compared with their Einstein radii that a non-binary lens is well 
described as a symmetric point-mass system.}  Every real 
cosmological lens will have some small asymmetries either in its 
own mass distribution (e.g., ellipticity), or in the distribution 
of objects near the line of sight (leading to a tidal shear).  In 
fact, it is well known that ellipticity and shear cannot be 
ignored in models of individual observed strong lens systems 
\citep*[e.g.,][]{KKS,witt}.

Nevertheless, most analyses of the statistics of gravitational 
lenses have used symmetric lenses.  The statistical calculations 
offer enough intrinsic challenges that most authors have stuck to 
idealized spherical lenses, such as the singular isothermal sphere 
(SIS) or the generalized Navarro-Frenk-White
\citep[GNFW;][]{NFW97,zhao} profile
\citep[e.g.,][]{narayan,fukugita,csk95,csk96a,maoz,KM01,sarbu,
takahashi,li,davis,HM04,KKM,mitchell}.
Conventional wisdom holds that the statistical effects of 
ellipticity and shear are confined to the relative numbers of 
double and quadruple lenses, and that symmetric lenses are 
adequate for applications such as deriving cosmological 
constraints.

To our knowledge, this conventional wisdom is based on a few 
studies in which the analysis of ellipticity and shear was 
subordinate to practical applications of lens statistics.  
\citet{king}, \citet{csk96b}, \citet{KKS}, and \citet{rusin2} all 
computed the relative abundances of different image configurations 
as a function of ellipticity and/or shear, for various assumptions 
about the luminosity function of background sources.  Along the 
way, they necessarily computed the effects of ellipticity and 
shear on the lensing cross section and magnification bias, but did 
not explicitly discuss them. \citet{chae} included ellipticity in 
lens statistics constraints on cosmological parameters, but the 
effects were built into his statistical machinery and not 
presented on their own.  We believe there is pedagogical value in 
isolating the statistical effects of ellipticity and shear and 
studying them in detail.  It would be useful to lay out exactly 
how ellipticity and shear affect the lensing optical depth, and 
how that may (or may not) lead to biases in cosmological 
constraints.  Furthermore, at least two other issues deserve to be 
studied as well: the effects of ellipticity and shear on the 
distribution of lens image separations; and the dependence of the 
various statistical effects on the luminosity function (LF) of the 
background sources.  We will show that, while not wrong, the 
conventional wisdom is somewhat limited and there are effects of
ellipticity and shear on lens statistics that are subtle but 
interesting.

We focus on lensing by galaxies, by which we mean systems with a 
single dominant mass component that can be approximated as an 
isothermal ellipsoid.  The isothermal profile describes early-type 
galaxies remarkably well on the 5--10~kpc scales relevant for
strong lensing 
\citep[e.g.,][]{rix97,gerhard,rusin,tk2016,kt1608,rkk}.
Lens statistics are rather different for groups and clusters of 
galaxies modeled with GNFW profiles, and that parallel case has 
recently been studied by \citet{oguri2}.

\section{Methodology}
\label{sec:method}

\subsection{General theory}
\label{sec:general}

The probability for a source at redshift $z_s$ to be lensed can be 
written as
\begin{eqnarray}
  \tau(z_s) &=&
    \frac{1}{4\pi}\int dV
    \int d\sigma\ \frac{dn}{d\sigma}
    \int de\ p_e(e)
    \int d^2\gamma\ p_\gamma(\gamma,\pg) \nonumber\\
  && \qquad \times \int_{\rm mult} d\u\ 
    \frac{\Phi_{\rm src}(L/\mu)}{\Phi_{\rm src}(L)}\ .
    \label{eq:tau}
\end{eqnarray}
The first integral is over the volume of the universe out to the 
source.  The second integral is over the population of galaxies 
that can act as deflectors.  For isothermal lenses the important 
physical parameter is the velocity dispersion, so the most useful 
description of the galaxy population is the velocity dispersion 
distribution function $(dn/d\sigma)d\sigma$, or the number density 
of galaxies with velocity dispersion between $\sigma$ and 
$\sigma+d\sigma$ \citep[see][]{mitchell}.  The third integral is 
over an appropriate distribution $p_e$ for the internal shape
(ellipticity) of the lens galaxy.  (Without loss of generality, we 
can work in coordinates aligned with the major axis of the galaxy 
so we do not need to consider the galaxy position angle.)  The 
fourth integral is over an appropriate distribution $p_\gamma$ for 
the external tidal shear caused by objects near the lens galaxy; 
this integral is two-dimensional because shear has both an 
amplitude ($\gamma$) and a direction ($\pg$).  Finally, the fifth 
integral is over the angular position $\u$ of the source in the 
source plane, and is limited to the multiply-imaged region.  In 
the last integrand, $\mu$ is the lensing magnification, $\Phi_{\rm 
src}(L)$ is the cumulative number density of sources brighter than 
luminosity $L$ in the survey, and the factor
$\Phi_{\rm src}(L/\mu)/\Phi_{\rm src}(L)$ accounts for the 
``magnification bias'' that produces an excess of faint sources in 
a flux-limited survey due to lensing magnification \citep{TOG}.  
(The role of the limiting flux or limiting luminosity will be 
discussed in \refsec{LF} below.)  The differential probability for 
having a lens with image separation $\dt$ can be computed by 
inserting a Dirac $\delta$-function in \refeq{tau} to select the 
parameter combinations that give separation $\dt$.  In other 
words, we can think of the (normalized) image separation 
distribution as $p(\dt) = \tau^{-1}\,\partial\tau/\partial\dt$.

The lensing cross section $A$ and the magnification bias factor 
$B$ are often computed separately (see \citealt{chae} for the most 
recent example).  According to \refeq{tau}, however, the two 
quantities are closely linked.  We prefer to keep them together 
and compute the product
\begin{equation} \label{eq:BA}
  BA \equiv \int_{\rm mult} d\u\ 
    \frac{\Phi_{\rm src}(L/\mu)}{\Phi_{\rm src}(L)}\ ,
\end{equation}
which we call the ``biased cross section.''  The biased cross 
section depends on both the lens model parameters and the source 
LF.

A convenient feature of isothermal lenses is that the physical 
scale decouples from the lensing properties.  All of the 
dependence on $z_s$, $z_l$, and $\sigma$ is contained in the 
(angular) Einstein radius $\theta_E$, so when we work in units of 
$\theta_E$ nothing depends explicitly on these parameters.  As a 
result, the dimensionless biased cross section
$\BA \equiv BA/\theta_E^2$ depends only on the ellipticity and 
shear (and implicitly on the source LF).  We can then rewrite 
\refeq{tau} as
\begin{eqnarray}
  \tau(z_s) &=&
    \left\{ \frac{1}{4\pi}\int dV
    \int d\sigma\ \frac{dn}{d\sigma}\ \theta^2_E(z_s,z_l,\sigma) 
\right\}
    \times \nonumber \\
  && \quad \left\{ \int de\ p_e(e)
    \int d^2\gamma\ p_\gamma(\gamma,\pg)\ \BA(e,\gamma) 
\right\}\,,
\end{eqnarray}
where the second factor depends only on the ellipticity and shear 
distributions, while the first factor depends only on the source 
redshift and the deflector population.  If we only want the
{\it change\/} in the optical depth produced by ellipticity and 
shear, then we can simply write\begin{equation} \label{eq:tau2}
  \frac{\tau}{\tau_0} = \int de\ p_e(e)
    \int d^2\gamma\ p_\gamma(\gamma,\pg)\
    \frac{\BA(e,\gamma,\pg)}{\BA_0}\ ,
\label{eq:tau_over_tau0}
\end{equation}
where $\BA_0$ and $\tau_0$ are the biased cross section and the 
optical depth for the spherical case.

Working in dimensionless units also simplifies the study of image 
separations.  Even if all galaxies are SIS, the distribution of 
image separations will be fairly broad because there is a range of 
lens galaxy masses and redshifts \citep[see, e.g.,][]{csk93a}.
However, the dimensionless separation $\that = \dt/\theta_E$ is 
always $\that=2$ for an SIS lens, so the dimensionless image 
separation distribution $p(\that)$ is just a $\delta$-function 
when the ellipticity and shear are zero.  This means that studying 
$p(\that)$ is the simplest way to identify {\it changes\/} to the 
image separation distribution caused by ellipticity and shear.
For fixed ellipticity and shear, the distribution can be formally 
written as
\begin{equation} \label{eq:sep}
  p(\that \mid e,\gamma,\pg) =
    \int_{\rm mult} d\u\ 
    \frac{\Phi_{\rm src}(L/\mu)}{\Phi_{\rm src}(L)}\,
    \delta\left[\that - \that(\u ; e,\gamma,\pg)\right]\,,
\end{equation}
where $\that(\u ; e,\gamma,\pg)$ is the dimensionless image 
separation produced for a source at position $\u$ by a lens with 
the specified ellipticity and shear.  The full image separation 
distribution can then be found by integrating over appropriate 
distributions of ellipticity and shear. Note that we do not 
actually need to integrate over the masses and redshifts of the 
deflector population in order to compute changes to the optical 
depth and the image separation distribution.

\subsection{The isothermal ellipsoid with shear}
\label{sec:SIEg}

We first discuss isothermal ellipsoids without an external shear, 
and then discuss properties of shear at the end of this 
subsection.

The projected surface mass density for an isothermal ellipsoid, 
written in polar coordinates $(r,\phi)$ and expressed in units of 
the critical density for lensing, is
\begin{equation} \label{eq:sie}
  \kappa(r,\phi) = \frac{\Sigma}{\Sigma_{\rm crit}}
  = \frac{b}{2r} \left[\frac{1+q^2}
    {(1+q^2)-(1-q^2)\cos2\phi}\right]^{1/2} ,
\end{equation}
where $q \le 1$ is the axis ratio, and the ellipticity is $e=1-q$.  
(Recall that we are working in coordinates aligned with the major 
axis of the galaxy.)  The radius $r$ and the parameter $b$ both 
have the dimensions of length, and may be expressed as physical 
lengths (e.g., kiloparsecs) or angles on the sky (radians or 
arcseconds); we work in angular units.  The lensing properties of 
an isothermal ellipsoid are given by \citet{kassiola},
\citet*{kormann}, and \citet{spirals}.

The Einstein radius sets the lensing scale, so it is useful to 
determine its value.  Consider the deflection $\alpha_0(r)$ 
produced by the monopole moment of the lens galaxy,
\begin{equation}
  \alpha_0(r) = \frac{1}{\pi r} \int_{0}^{r} dr'
    \int_{0}^{2\pi} d\phi\ r'\ \kappa(r',\phi)
  = \frac{M_{\rm cyl}(r)}{\pi\,r\,\Sigma_{\rm crit}}\ ,
    \label{eq:def0}
\end{equation}
where $M_{\rm cyl}(r)$ is the projected mass in a cylinder of 
radius $r$.  The Einstein radius is defined by
$\alpha_0(\theta_E) = \theta_E$.  This definition reduces to the 
standard Einstein radius in the spherical case, and it is the 
quantity that seems to be most relevant in models of nonspherical 
lenses \citep[e.g.,][]{cohn}.  For the isothermal ellipsoid, we 
find
\begin{equation}
  \frac{\theta_E}{b} = \frac{1}{\pi}\
    \left[2(1+q^{-2})\right]^{1/2}\ K\left(1+q^{-2}\right)\,,
\end{equation}
where $K(x)$ is the elliptic integral of the first kind.  For
reference we note that this function can be approximated by
\begin{equation}
  \frac{\theta_E(e)}{b} = \exp{\left[(0.89e)^3\right]}\,,
\end{equation}
which is accurate to $<$1\% for $e \leq 0.53$ and to $<$4\% for
$e \leq 0.9$.  In practice, however, we use the exact result.

\begin{figure}
\centerline{
\epsfig{file=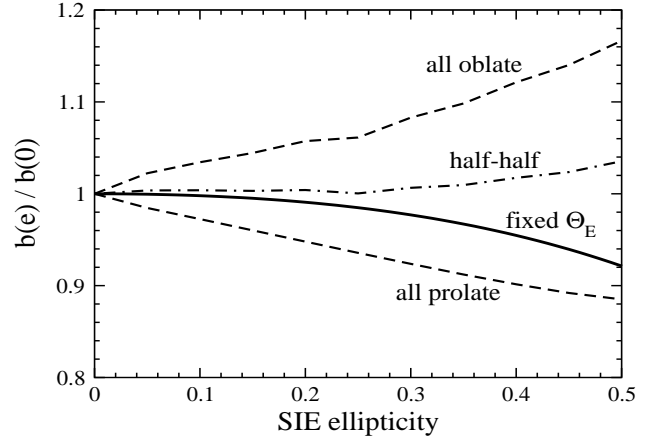, height=3.5in, width=2.5in, angle=-90}
}
\caption{
Change in the isothermal ellipsoid $b$ parameter as a function of 
ellipticity.  The two dashed curves show the dynamical 
normalization if all halos are assumed to be oblate or prolate.
The dot-dashed curve shows the case when half of the halos are
assumed to be oblate and half prolate.  The solid curve shows the
result when the Einstein radius is fixed to be independent of
ellipticity, which is what we assume in this paper.
}\label{fig:norm}
\end{figure}

We must specify how to normalize the model, or how to choose the 
parameter $b$.  For a spherical galaxy, $b$ simply equals the 
Einstein radius and is related to the velocity dispersion by
\begin{equation} \label{eq:SISb}
  b = \theta_E = 4\pi \left(\frac{\sigma}{c}\right)^2\ 
\frac{D_{ls}}{D_{os}}\ ,
\end{equation}
where $D_{os}$ and $D_{ls}$ are angular diameter distances from 
the observer to the source and from the lens to the source.  For a 
nonspherical galaxy the situation is less straightforward.  If we 
seek a dynamical normalization in terms of a measurable stellar 
velocity dispersion, then we must worry about complications 
involving the halo shape and projection effects 
\citep{KKS,spirals,chae}.  Consider the dynamical normalization 
shown in \reffig{norm} \citep[following][]{chae}.  At a typical 
ellipticity $e \approx 0.3$, $b$ could rise by 7\% (compared to 
the spherical value) if all halos are oblate, or fall by 7\% if 
all halos are prolate.  Some dissipationless numerical simulations
have predicted roughly comparable numbers of oblate and prolate
halos \citep{dubinski,jing_suto}, which would yield a $b$ value
less than 1\% higher than the spherical value (for $e=0.3$).
However, the shape distribution is likely to be affected by
hydrodynamics \citep[e.g.,][]{stelios}, so it is not understood
in detail (in simulations, let alone in reality).  In other words,
the dynamical normalization appears to be small but uncertain, and
impossible to compute precisely.

An alternate approach is to fix the Einstein radius to be 
independent of ellipticity (and shear; see below).  This seems 
reasonable, because the Einstein radii of observed lenses can 
generally be determined in a model-independent way to a few 
percent accuracy \citep*[e.g.,][]{cohn,munoz}, and because it 
keeps the mass properties (the aperture mass) independent of 
ellipticity and shear.  This normalization is also shown in 
\reffig{norm}, and it is the one we adopt.  However, it is 
important to keep in mind that there is an irreducible uncertainty 
of a few percent in our analysis associated with the normalization.

Objects in the vicinity of the lens galaxy create tidal forces 
that affect the lens potential.  The contribution is often modeled 
as an external shear whose contribution to the lens potential is
\begin{equation}
  \Phi_{\rm shear}(r,\phi) = - \frac{\gamma}{2}\,r^2\,\cos2(\phi-
\pg)\,,
\end{equation}
where $(r,\phi)$ are polar coordinates, $\gamma$ is the 
dimensionless shear amplitude, and $\pg$ is the direction of the 
shear.  As an example, consider the shear produced by an 
isothermal sphere galaxy with Einstein radius $b_0$ that lies at 
polar coordinates $(r_0,\phi_0)$ relative to the lens galaxy; the 
shear amplitude would be $\gamma = b_0/(2r_0)$ and the shear 
direction would be $\pg = \phi_0$.  External shear does not 
contribute to the local surface mass density, so it does not 
affect the monopole deflection or the Einstein radius.

\subsection{Source luminosity functions}
\label{sec:LF}

The number density of sources with luminosity between $L$ and 
$L+dL$ is given by the luminosity function
$[d\phi_{\rm src}(L)/dL]\,dL$.  The quantity of interest for lens 
statistics (see \refeq{tau}) is the cumulative number density of 
sources brighter than $L$, or
\begin{equation} \label{eq:Phi}
  \Phi_{\rm src}(L) = \int_{L}^{\infty} 
  {d\phi_{\rm src}(L')\over dL' }\,dL'\,.
\end{equation}
We consider model LFs appropriate to both radio and optical 
surveys.

The simplest model LF is a featureless power law,
$\phi_{\rm src}(L) \propto L^{-\beta}$.  In this case the biased 
cross section simplifies to
\begin{equation} \label{eq:BAplaw}
  BA = \int_{\rm mult} \mu^{\beta-1}\ d\u
  \equiv \int \mu^{\beta-1}\ p(\mu)\ d\mu\,,
\end{equation}
where $p(\mu)$ is the distribution of magnifications for lensed 
sources.  Several points are worth mentioning.  First, with
$\beta \to 1^{+}$ the magnification weighting factor becomes
unity and we recover the simple lensing cross section with no 
magnification bias.\footnote{We write $\beta \to 1^{+}$, meaning 
that $\beta$ approaches unity from above, because for
$\beta \le 1$ the cumulative LF integral (\refeq{Phi}) formally 
diverges.  Nevertheless, the biased cross section integral 
(\refeq{BAplaw}) remains well defined.}  Second, because the 
magnification distribution generically has a power law tail 
$p(\mu) \propto \mu^{-3}$ for $\mu \gg 1$ \citep*[see][]{SEF}, the 
integral diverges for $\beta \ge 3$ and the biased cross section 
is well defined only for $\beta < 3$.  Finally, a power law is 
featureless so the biased cross section does not depend on the 
particular flux or luminosity limit of a survey.  A power law LF 
is a good model for radio surveys.  For example, the largest 
existing lens survey is the JVAS/CLASS survey of flat-spectrum 
radio sources \citep{class1,class2}, and it has an LF that is well 
described by a power law with $\beta \approx 2.1$ 
\citep[see][]{rusin2,chae}.

Future lens samples are likely to be dominated by
optically-selected quasar lenses found in deep wide-field imaging 
surveys \citep[e.g.,][]{KKM}.  While accurate determination of the 
quasar LF is a long-standing problem, recent evidence favors the 
double power law form proposed by \citet*{boyle},
\begin{equation}
  {d\phi(L,z)\over  dL}\,dL = \frac{\phi_*}
    {[L/L_*(z)]^{\beta_l} + [L/L_*(z)]^{\beta_h}}\ 
\frac{dL}{L_*(z)}\ ,
\end{equation}
where the break luminosity $L_*$ evolves with redshift as
\citep{madau99}
\begin{equation}
  L_*(z) = L_*(0)\ (1+z)^{\alpha_{\rm s}-1}\ 
   \frac{e^{\zeta z}(1+e^{\xi z_*})}{e^{\xi z}+e^{\xi z_*}}\ ,
\end{equation}
where the quasar spectral energy distribution is assumed to be a 
power law, $f_\nu\propto \nu^{-\alpha_{\rm s}}$.  With this LF, 
the biased cross section clearly depends on the bright and faint 
slopes $\beta_h$ and $\beta_l$, and also on the limiting 
luminosity $\Lcut(z)/L_*(z)$.  It depends on source redshift to 
the extent that these quantities depend on redshift.  We adopt the 
model from \citet{fan} with bright-end slope $\beta_h=3.43$ at 
$z<3$ and $\beta_h=2.58$ at $z>3$, and faint-end slope 
$\beta_l=1.64$ at all redshifts \citep{wyithe}.

If we wanted to compute statistics for real quasar lens surveys, 
we would need to adopt an appropriate limiting magnitude and 
compute the limiting luminosity $\Lcut(z)/L_*(z)$ as a function of 
redshift.  This would require specifying the passband, computing 
$K$-corrections, and other details that would muddy the waters.  
Since the goal is conceptual understanding of the effects, we 
believe that it is simpler and more instructive to work with a 
luminosity cut $\Lcut/L_*$.  In this case, we do not need to 
specify the parameters $\phi_*$, $L_*(0)$, $z_*$, $\zeta$, and 
$\xi$.

\subsection{Numerical techniques}
\label{sec:monte}

We compute the integrals in eqs.~(\ref{eq:BA})--(\ref{eq:sep}) 
using Monte Carlo techniques.  First, for fixed ellipticity and 
shear we place $10^5$--$10^6$ random sources in the source plane, 
in the smallest circle enclosing the caustics.  We solve the lens 
equation using the {\it gravlens\/} software \citep{keeton} to 
determine the number of images and their positions and 
magnifications.  We define the image separation to be the maximal 
separation between any two images in the system,
$\Delta\theta\equiv \max |\vec{\theta_i}-\vec{\theta_j}|$; this is 
a convenient, observable, and well-defined quantity that is 
independent of the number of images.

We separate the lenses into three standard classes based on the 
image multiplicity: ``doubles'' have two bright images, one with 
positive parity and one negative, plus a faint central image that 
is rarely observed; ``quads'' have four bright images, two 
positive and two negative parity, plus a faint central image that 
is rarely observed; and ``naked cusps'' have three bright images, 
either two positive parity and one negative or vice versa.  We use 
the classifications directly only when studying the quadruple-to-
double ratio (\refsec{qd}).  The classification offers a fringe 
benefit: we can identify numerical errors as systems that do not 
fit into any of the classes (because, for example, the software 
failed to find one of the images).  We estimate that the numerical 
failure rate is $<\!10^{-4}$.

Next, where appropriate we integrate over distributions for 
ellipticity and shear.  For the ellipticity, we adopt the 
distribution of ellipticities measured for 379 early-type galaxies 
in 11 nearby clusters by \citet{jorgensen}.  The distribution has 
mean $\avg{e} = 0.31$ and dispersion $\sigma_e = 0.18$, and there 
are no galaxies with $e \gtrsim 0.8$.  Although the measured 
ellipticities describe the luminosity while what we need for 
lensing is the ellipticity of the mass distribution, this is 
probably the best we can do at the moment.  In any case, it seems 
reasonable to think that the ellipticity distributions for the 
light and the mass may be similar \citep[see][]{rusin2}.  For the 
shear, \citet{holder} estimate that the distribution of shear 
amplitudes derived from simulations of galaxy formation can be 
described as a lognormal distribution with median $\gamma = 0.05$ 
and dispersion $\sigma_\gamma = 0.2$ dex; this distribution is 
also broadly consistent with the shears required to fit observed 
lenses.  As a rule of thumb, a shear $\gamma \sim 0.1$ is common 
for lenses in poor groups of galaxies, and the shear can reach 
$\gamma \sim 0.3$ for lenses in rich clusters 
\citep[e.g.,][]{KKS,kundic1115,kundic1422,fischer,kneib}.  We 
assume random shear orientations.

\section{The Optical Depth}
\label{sec:tau}

\begin{figure}
\centerline{
\epsfig{file=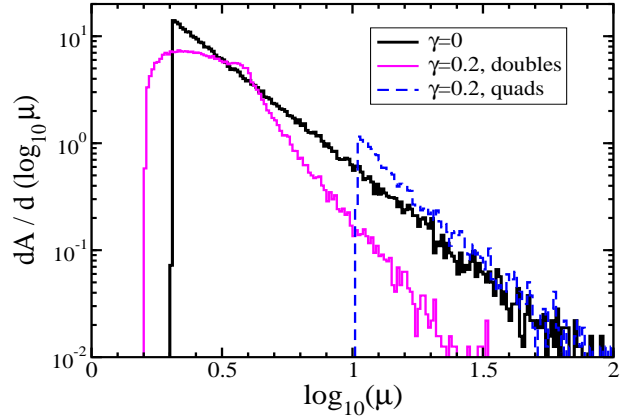, height=3.5in, width=2.5in, angle=-90}
}
\caption{ 
Magnification distributions for spherical deflectors without shear 
and with shear $\gamma=0.2$. The curves are normalized so that the 
area under each curve is the corresponding cross section in units 
of $\theta_E^2$.  For the shear case we show the distributions for 
doubles and quadruples separately.  Note that the minimum 
magnification for doubles is
$\mu_{2, \rm min} = 2/[(1+3\gamma)(1-\gamma)]$, while for 
quadruples it is $\mu_{4, \rm min} = 2/[\gamma(1-\gamma^2)]$
\citep{finch}; so $\mu_{2, \rm min} = 1.56$ and
$\mu_{4, \rm min} = 10.4$ for the case $\gamma=0.2$ shown here. 
The distributions asymptote to $A(\mu) \propto \mu^{-3}$ at high 
magnifications \citep{SEF}.
}\label{fig:magdist}
\end{figure}

Before determining the effects of ellipticity and shear on the 
lensing optical depth, it is instructive to consider first how 
they affect the source plane.  There is only a small change in the 
lensing cross section.  In fact, for isothermal galaxies shear has 
{\it no effect\/} on the radial caustic and hence on the cross 
section.\footnote{Shear can affect the cross section only in the 
rare case that the tangential caustic pierces the radial caustic 
to form naked cusps \citep[e.g.,][]{SEF}.  For SIS+shear models, 
this happens only when the shear is large, $\gamma>1/3$.  In this 
case, there is some (small) multiply-imaged region outside the 
radial caustic.}  Ellipticity (or any other internal angular 
structure) in isothermal galaxies changes the caustics in such a 
way as to {\it reduce\/} the cross section, as explained in the 
Appendix.  The main effect of increasing ellipticity or shear is 
to lengthen the tangential caustic, which enlarges the phase space 
for large magnifications and raises the tail of the magnification 
distribution, as illustrated in \reffig{magdist}.  In particular, 
we see a sharp increase in the cross section for producing 
magnifications larger than the minimum magnification for a 
quadruple lens (see caption).

\begin{figure*}
\centerline{
\epsfig{file=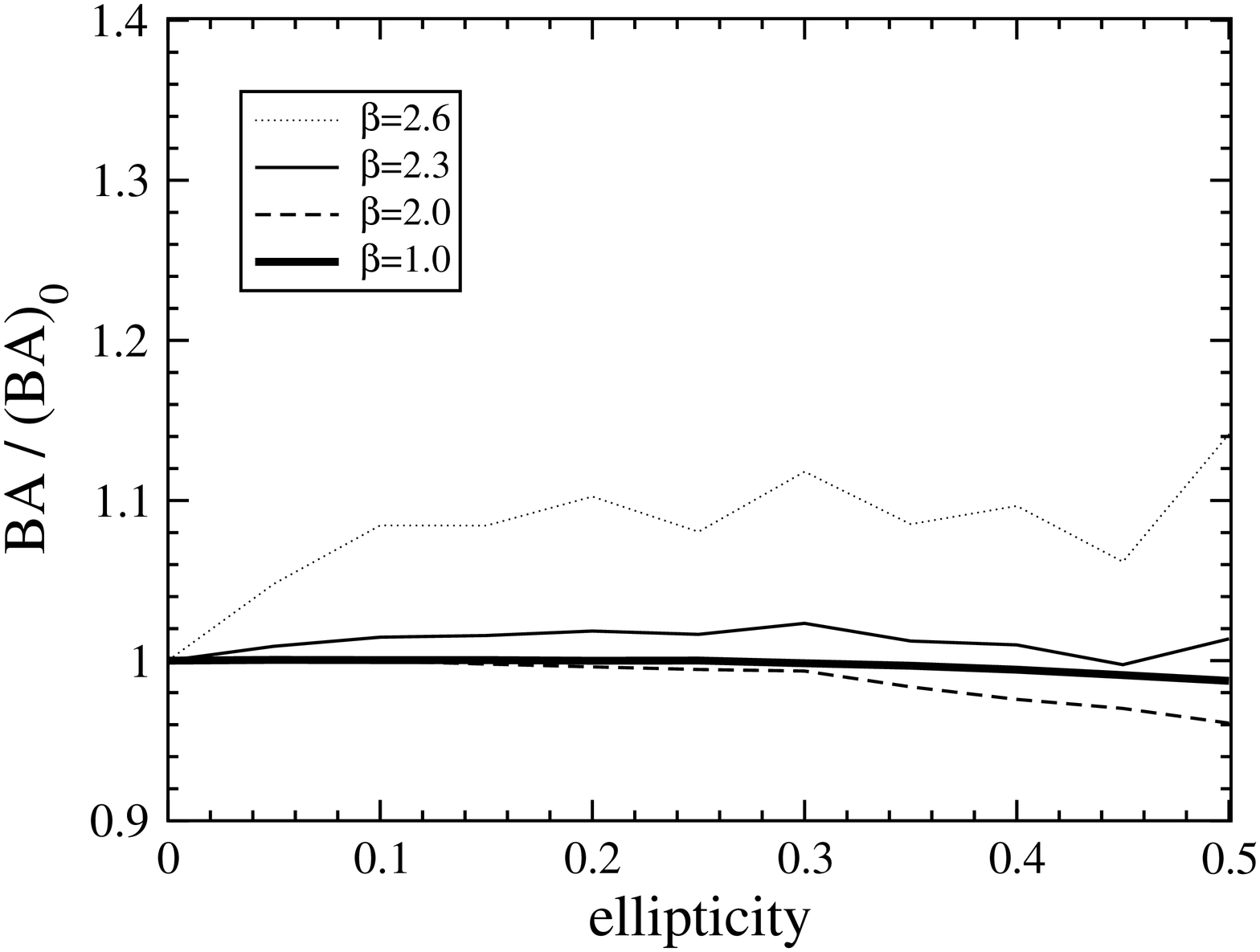, height=3.5in, width=2.5in, 
angle=-90}
\epsfig{file=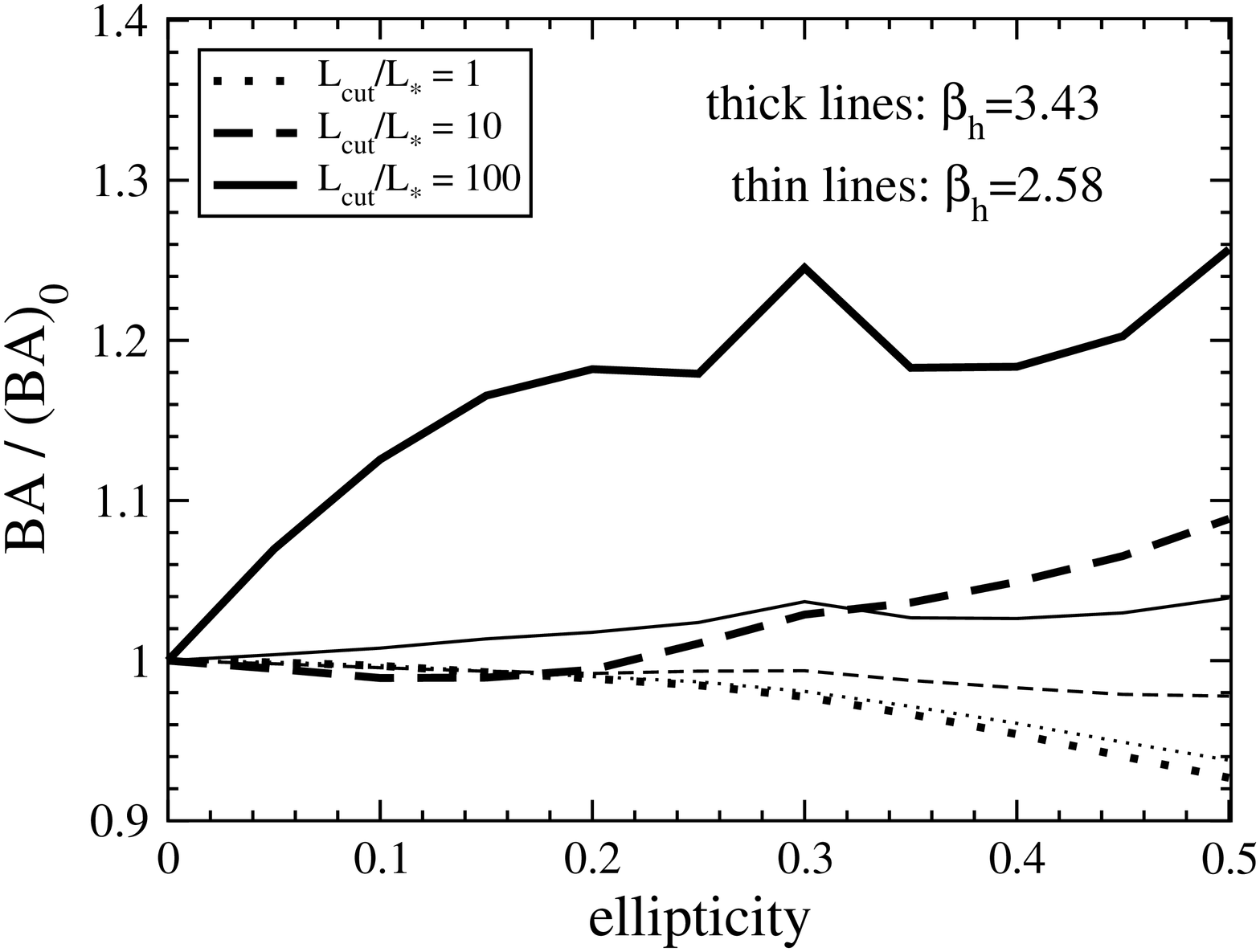, height=3.5in, width=2.5in, angle=-90}
}
\caption{
Enhancement in the biased cross section caused by ellipticity.  
The shear is set to zero.
(a, left) Results for different power law source LFs.  Recall that 
with $\beta \to 1^{+}$ there is no magnification bias, and that 
the CLASS radio survey has $\beta \approx 2.1$.
(b, right) Results for the model quasar LF, for different values 
of the limiting luminosity ($\Lcut/L_*$) and the bright-end slope 
($\beta_h$); the faint-end slope is fixed at $\beta_l=1.64$.  The 
jaggedness in the upper curves is due to statistical noise, 
because for steep LFs the magnification bias is strong and the 
results are dominated by rare extreme-magnification systems.  We 
estimate the statistical errors in the upper curves to be 
$\sim$3\%, and much smaller for the other curves.
}\label{fig:BA-e}
\end{figure*}

\begin{figure*}
\centerline{
\epsfig{file=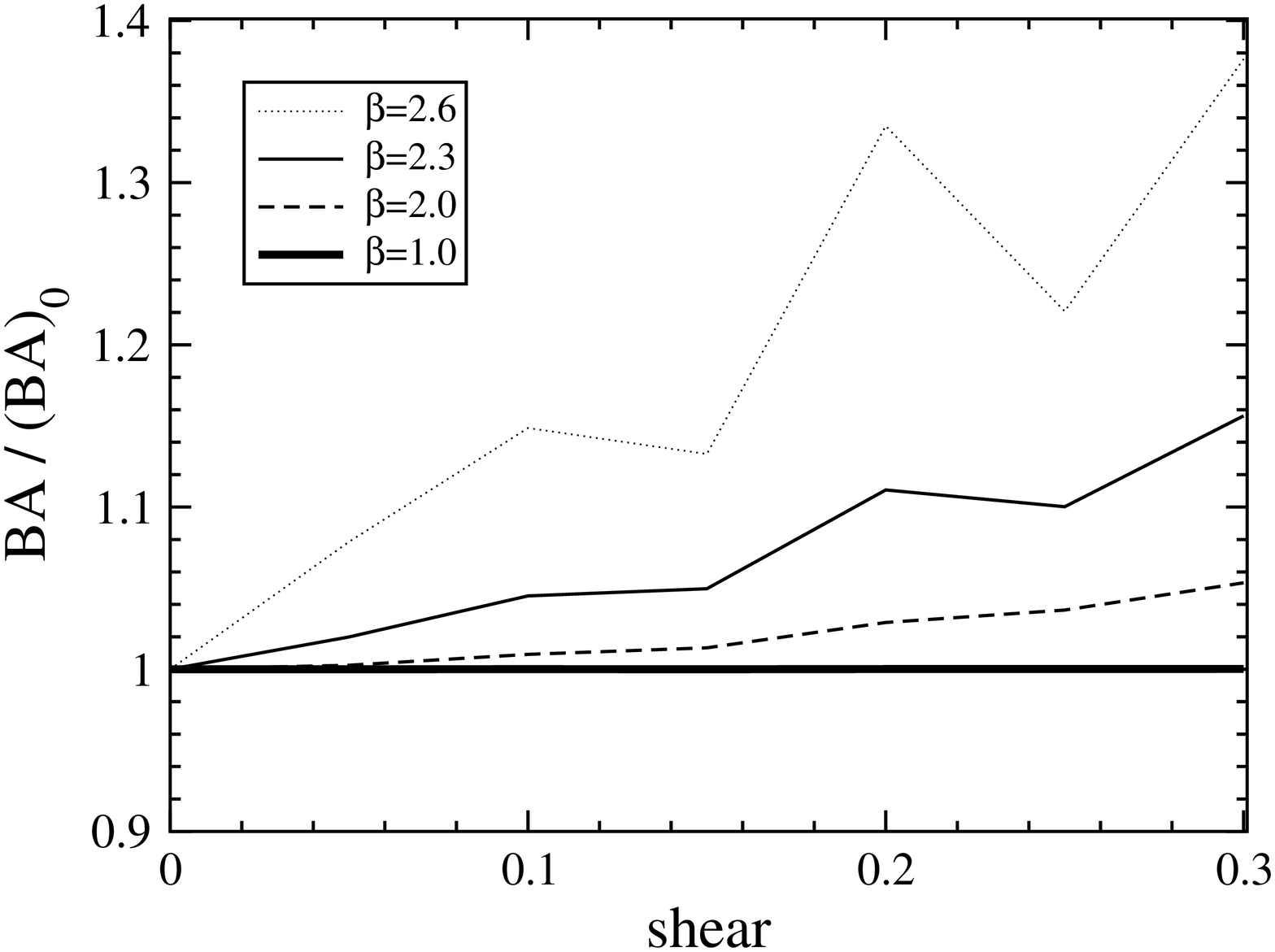, height=3.5in, width=2.5in, 
angle=-90}
\epsfig{file=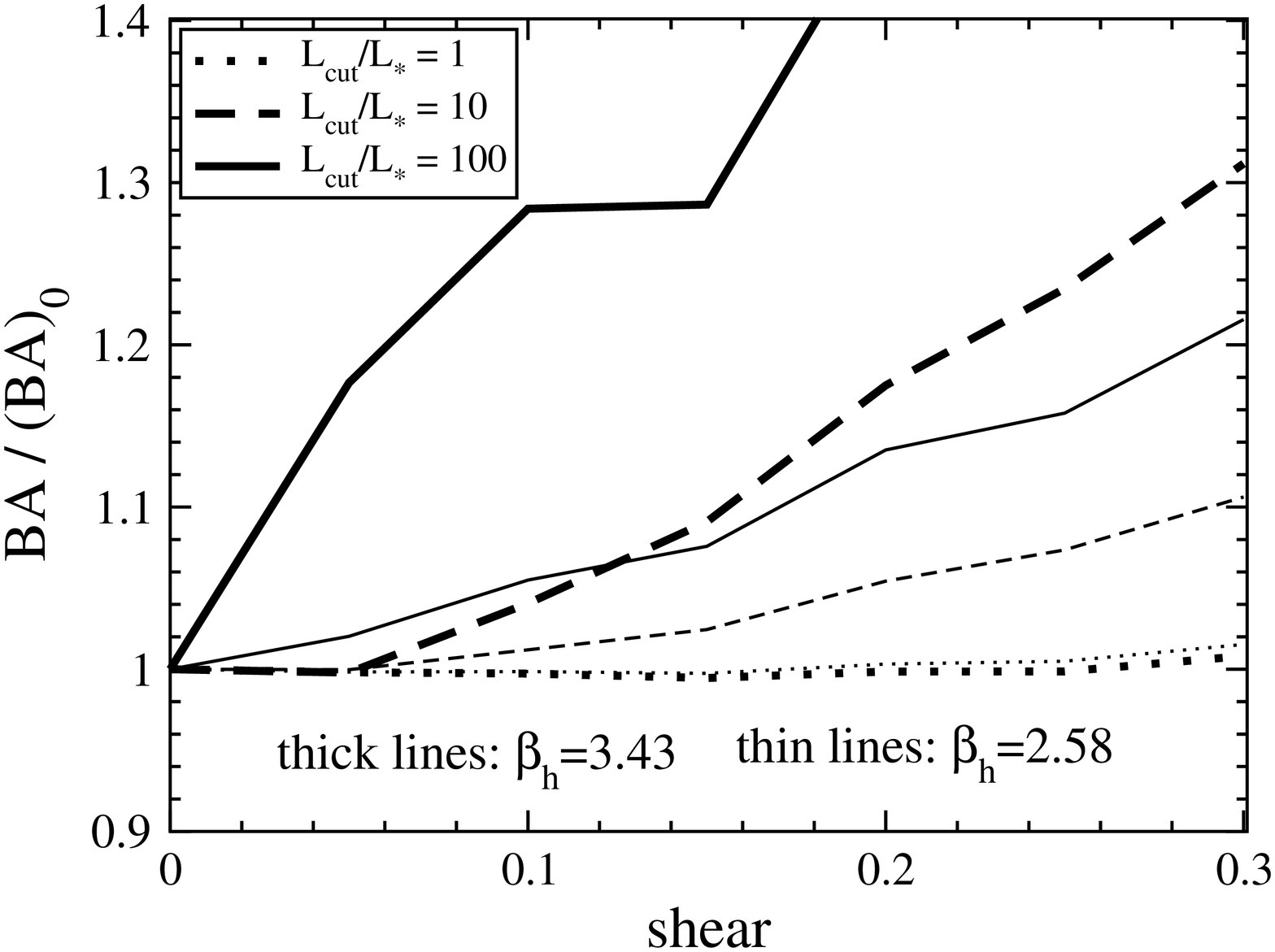, height=3.5in, width=2.5in, angle=-90}
}
\caption{
Similar to \reffig{BA-e}, but for shear.  The ellipticity is set 
to zero here.  The statistical errors in the upper curves are 
$\sim$10\%, and much smaller for the other curves.
}
\label{fig:BA-g}
\end{figure*}

We now examine the dependence of the biased cross section $BA$ on 
ellipticity (\reffig{BA-e}) and shear (\reffig{BA-g}).  When the 
LF is a power law with $\beta \to 1^{+}$ there is no magnification 
bias, and \reffig{BA-e} illustrates how ellipticity reduces the 
cross section.  Even with magnification bias, ellipticities up to 
$e \sim 0.5$ do not affect the biased cross section by more than 
10\% unless the source LF is very steep (e.g., the very brightest 
quasars, $\Lcut/L_* \gtrsim 100$ when $\beta_h = 3.43$).  
\reffig{BA-g} shows that shear causes a stronger increase in the 
biased cross section, but we must remember that realistic shears 
are $\gamma \lesssim 0.1$ and only lenses in clusters experience 
large shears of $\gamma \sim 0.2$--0.3.  Thus, the typical change 
in the biased cross section due to shear is again no more than 
10\% unless the LF is very steep.

\begin{figure*}
\centerline{
\epsfig{file=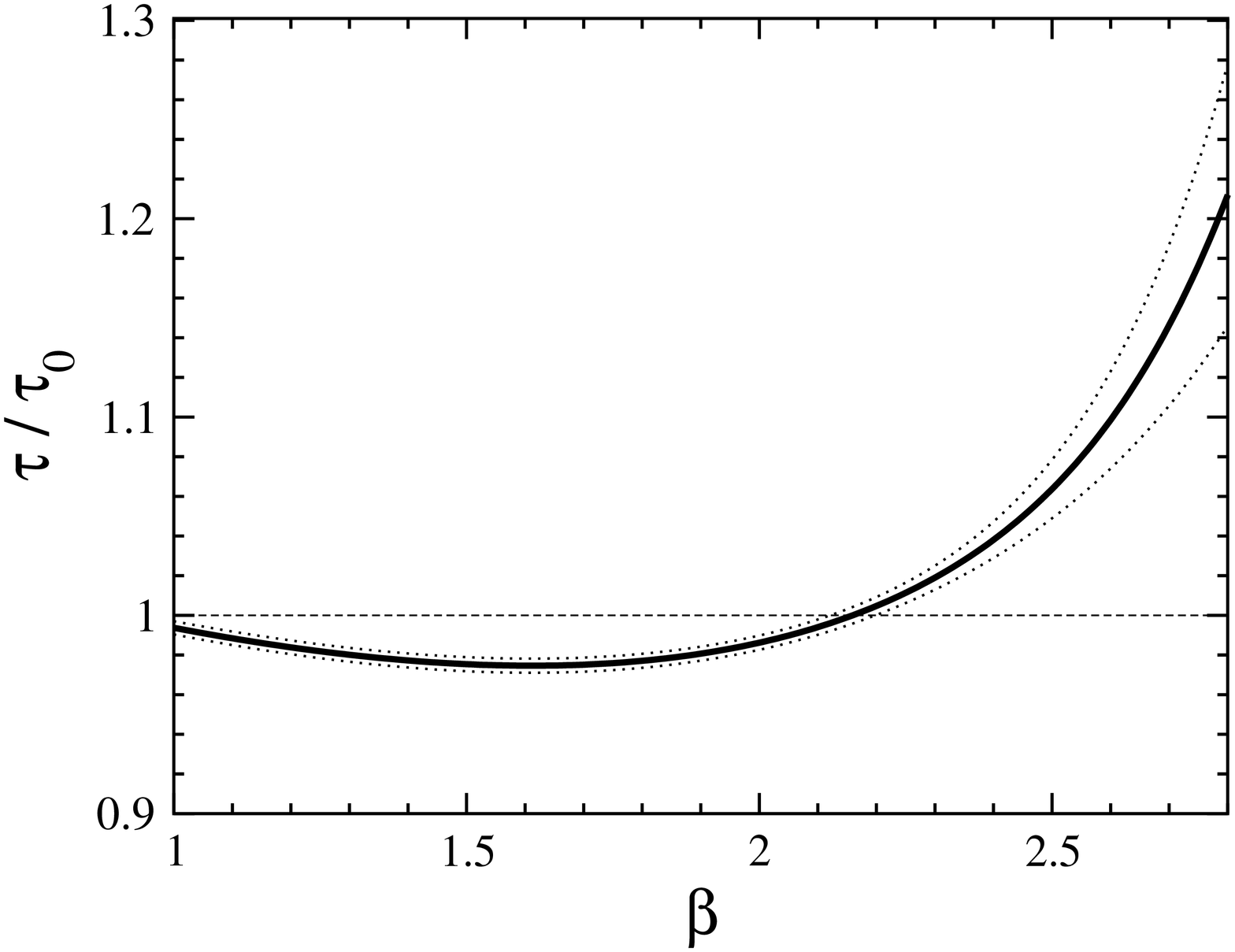, height=3.5in, width=2.5in, angle=-90}
\epsfig{file=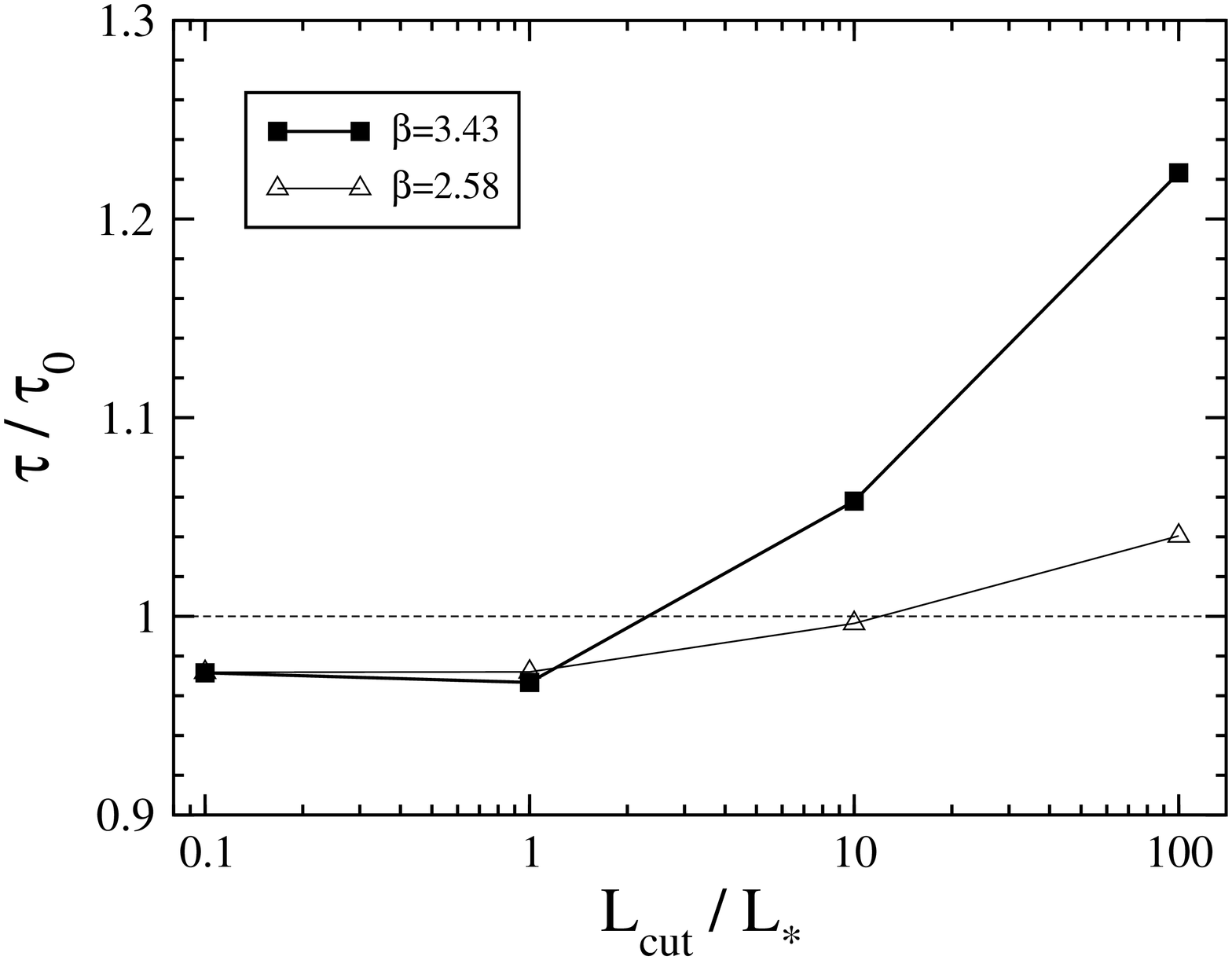, height=3.5in, width=2.5in, angle=-90}
}
\caption{
Enhancement in the optical depth as a function of the luminosity 
function slope $\beta$ for power law LFs (left) or the limiting 
luminosity $\Lcut/L_*$ for quasar LFs (right). In the left panel, 
the solid line shows the fiducial result while the dotted lines 
indicate the statistical uncertainty from our calculation.
}\label{fig:tauavg}
\end{figure*}

To compute changes in the full optical depth we integrate the 
biased optical depth over appropriate distributions of ellipticity 
and shear (see \refeq{tau_over_tau0}).  The results are shown in 
\reffig{tauavg}.  Without magnification bias ($\beta \to 1^{+}$), 
ellipticity and shear reduce the optical depth very slightly (by 
$\sim$0.6\%, although the statistical uncertainty from our Monte 
Carlo calculations is $\sim$0.3\%)  With magnification bias and a 
shallow source LF, ellipticity and shear can reduce the optical 
depth by up to $\sim$2.5\% relative to the spherical case.  For 
power law LFs, only when $\beta \ge 2.2$ is there an increase in 
the optical depth, and we must have $\beta \ge 2.5$
($\beta \ge 2.6$) in order for the increase to be more than 5\% 
(10\%).  For the quasar LF, the increase exceeds 5\% only if the 
bright end is steep ($\beta_h=3.43$) and the survey is limited to 
bright quasars ($\Lcut/L_* \gtrsim 10$).

In practice, these results mean that ellipticity and shear are 
important for the optical depth only in surveys that are 
restricted to the brightest quasars.  They are not very 
significant for the sorts of deep optical surveys now underway 
that probe well beyond the break in the quasar LF.

\section{The Image Separation Distribution}
\label{sec:sep}

We now turn to the distribution of lens image separations and how 
it is affected by ellipticity and shear.  First, we recall several 
basic facts.  Even in spherical models the distribution of 
dimensioned image separations will have some natural spread 
because of the range of galaxy masses and redshifts; but the 
distribution of dimensionless separations $\that$ is a
$\delta$-function at $\that=2$.  To highlight {\it changes\/} in 
the separation distribution, it is therefore useful to focus on 
the distribution of $\that$.  Also, as discussed in 
\refsec{monte}, we define the separation to be the maximal 
distance between any pair of images.

\begin{figure}
\centerline{
\epsfig{file=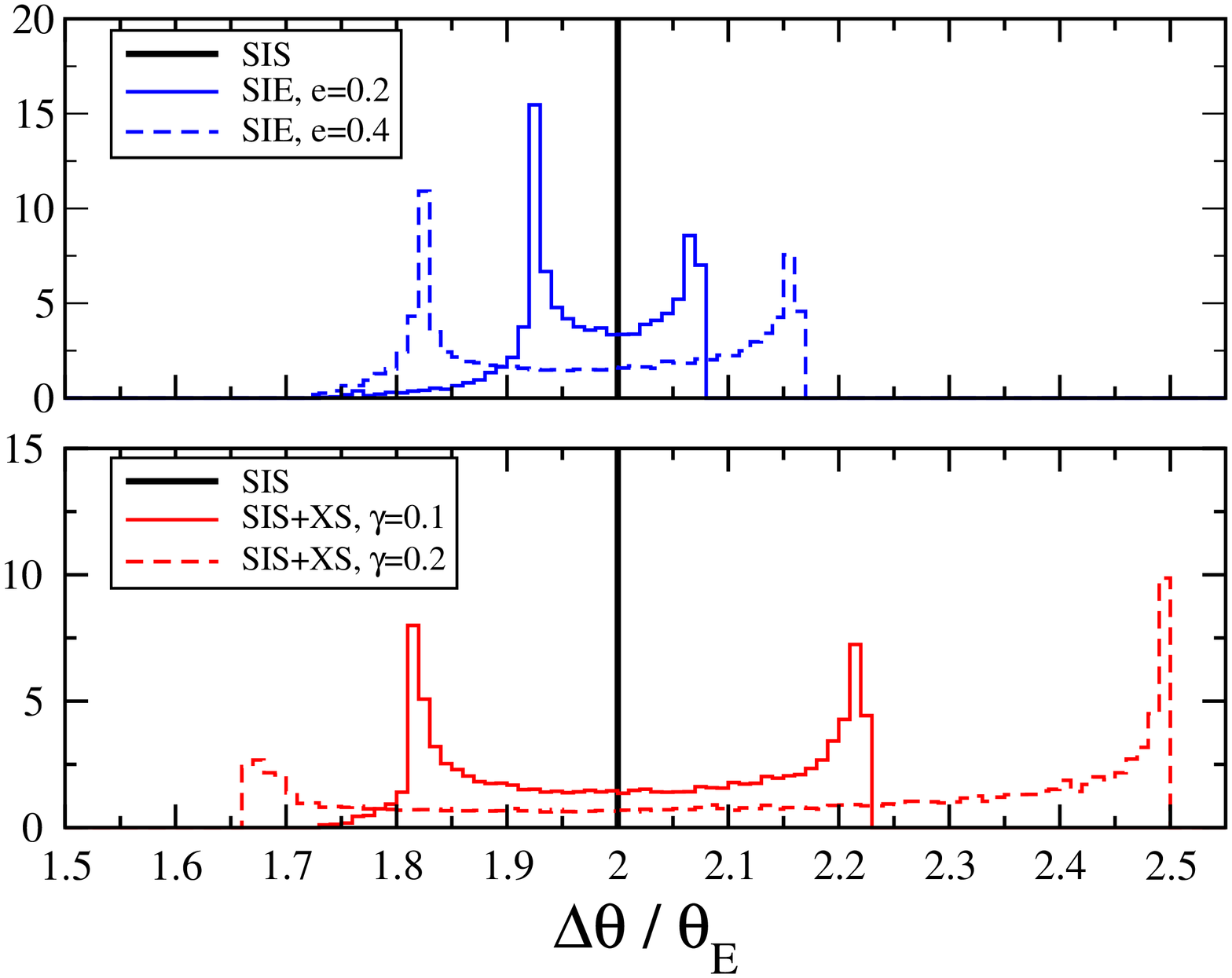, height=3.5in, width=2.5in, 
angle=-90}
}
\caption{
Histograms of the dimensionless image separation 
$\that=\dt/\theta_E$, for two values of ellipticity (top panel) 
and shear (bottom). For a spherical lens the distribution is a 
$\delta$-function at $\that=2$, indicated by the vertical line.  
The histograms contain all image multiplicities (i.e., both 
doubles and quadruples).  The results are shown for the CLASS LF 
(a power law with $\beta=2.1$), but they are not very sensitive to 
this choice.
}\label{fig:sep1}
\end{figure}

\reffig{sep1} shows the distribution of $\that$ for several values 
of ellipticity (upper panel) and shear (lower panel).  The 
distribution has an interesting shape that peaks at the ends and 
is low in the middle.  It has a sharp cutoff at the high end, 
while at the low end it has a sharp drop followed by a small tail 
to lower values.  The peaks correspond to sources near the minor 
and major axes of the lens potential.

\begin{figure}
\centerline{\epsfig{file=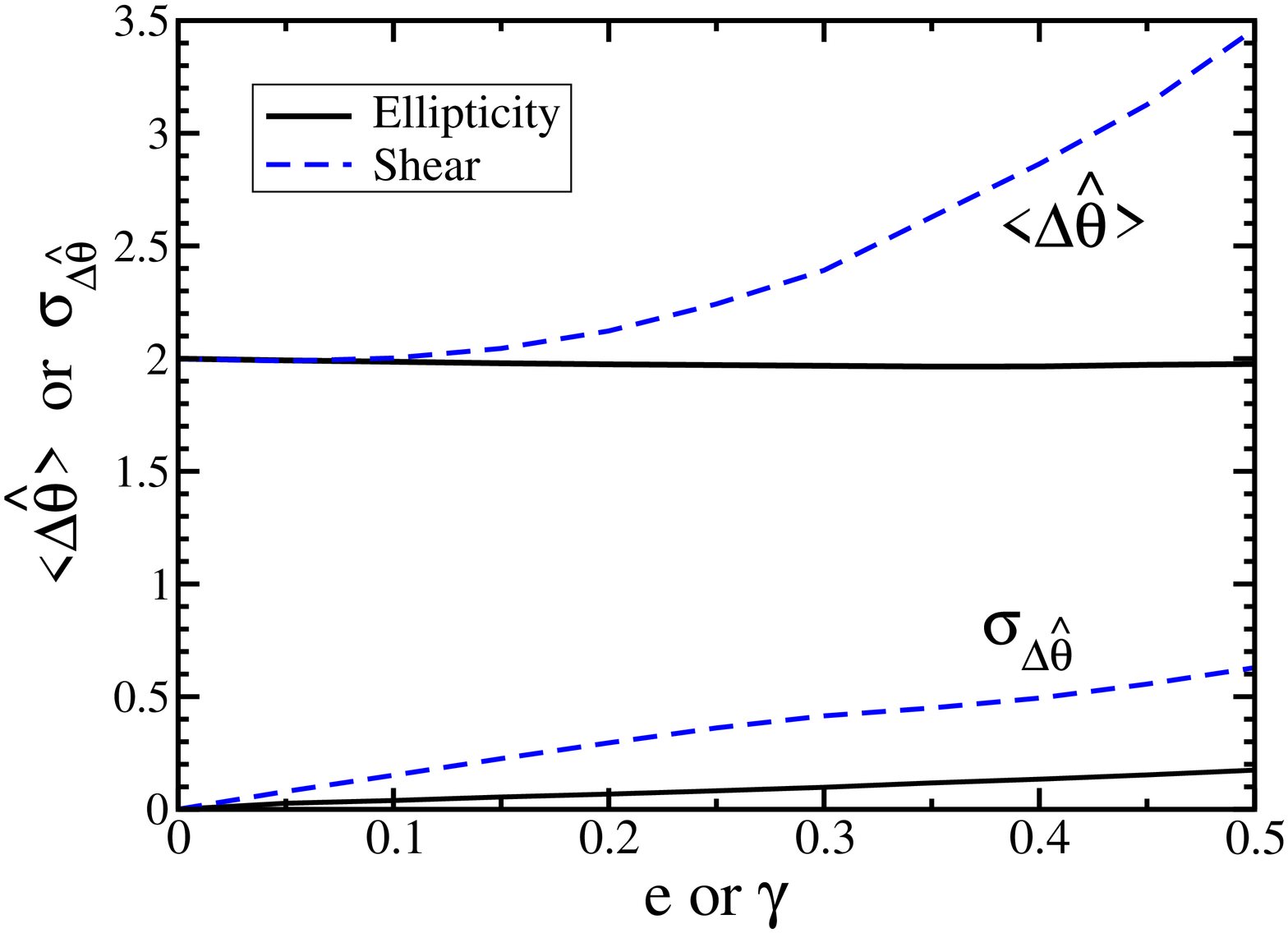, height=3.5in, 
width=2.5in, 
angle=-90}}
\caption{
Mean and spread of the image separation distribution as a function 
of ellipticity or shear.  The image separations are in units of 
$\theta_E$.  As in \reffig{sep1}, the results are shown for the 
CLASS LF but are not very sensitive to this choice.
}\label{fig:sep2}
\end{figure}

As the ellipticity or shear increases, the distribution of $\that$ 
broadens and its mean shifts.  To quantify these effects, we 
compute the mean separation $\avg{\that}$ and the spread 
$\sigma_{\that} = (\avg{\that^2}-\avg{\that}^2)^{1/2}$, and plot 
them as a function of ellipticity or shear in \reffig{sep2}.  The 
increase in the mean and scatter are small for all ellipticities, 
and are both $<$20\% for all but the strongest shears
($\gamma \gtrsim 0.3$) felt by lenses in cluster environments.  
Nevertheless, it is interesting that shear produces a net
{\it bias\/} toward larger image separations.

\begin{figure}
\centerline{
\epsfig{file=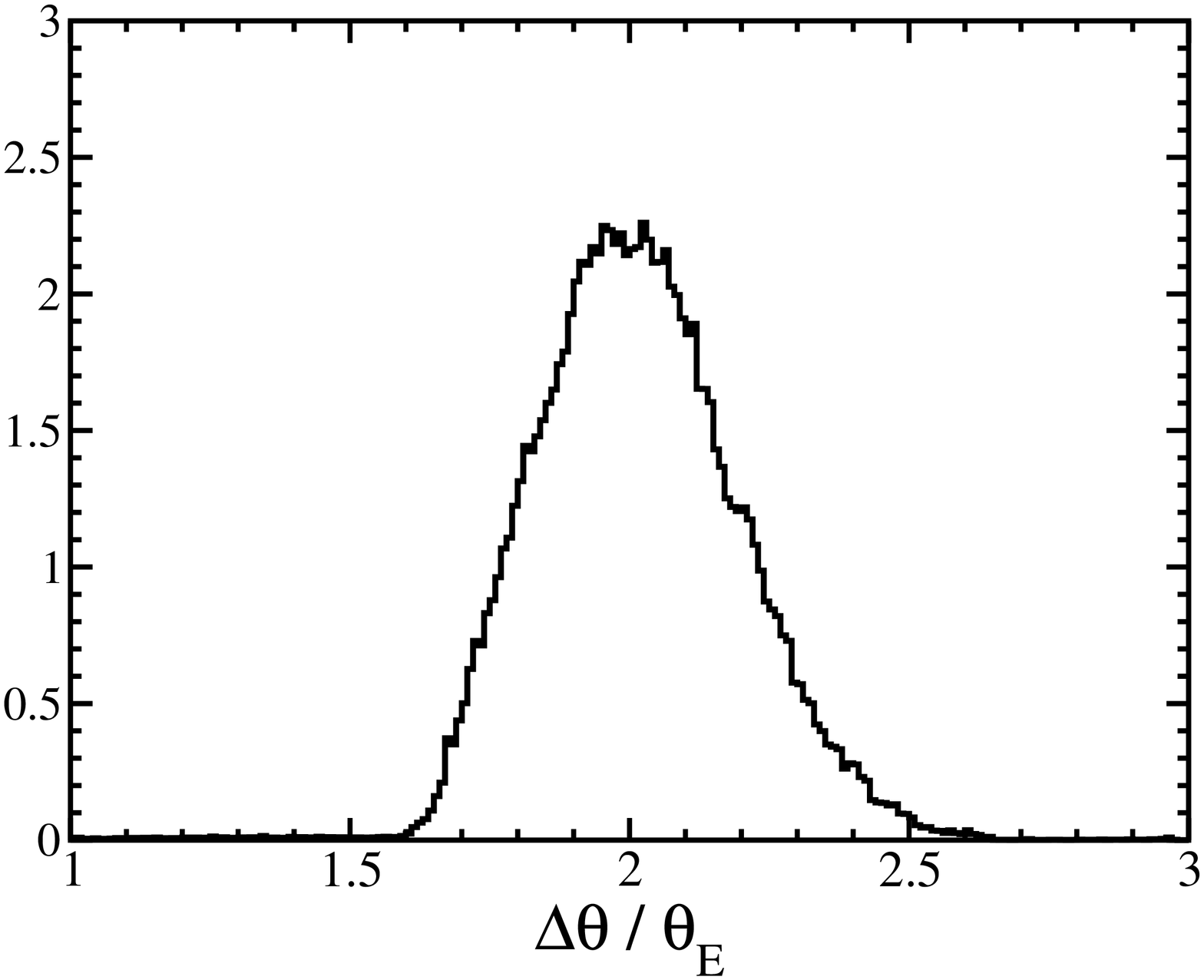, height=3.5in, width=2.5in, 
angle=-90}
}
\caption{
Net image separation distribution after averaging over ellipticity 
and shear.  The histogram is normalized to unit area.  The results 
are again shown for the CLASS LF but are not very sensitive to 
this choice. The distribution is nearly Gaussian, with mean
$\that = 2.01$ and scatter $\sigma_{\that} = 0.18$.
}\label{fig:sep3}
\end{figure}

Finally, by averaging over the ellipticity and shear distributions 
we obtain the net image separation distribution shown in 
\reffig{sep3}.  The averaging has smoothed out the sharp features 
seen in \reffig{sep1} when the ellipticity and shear were fixed.  
The net distribution is nearly Gaussian, with mean $\that = 2.01$ 
and scatter $\sigma_{\that} = 0.18$ for a power law LF with
$\beta = 2.1$, or $\that \approx 2.01$ and
$\sigma_{\that} \approx 0.19$ for various cases of the quasar LF.  
In other words, ellipticity and shear basically leave the mean 
image separation unchanged but create an additional scatter of 
10\%, and these results are insensitive to the source LF.

\section{Quadruple-to-Double Ratio}
\label{sec:qd}

We next consider how ellipticity and shear affect the number of 
lenses with different image configurations.  While an SIS lens 
always produces two images, increasing ellipticity or shear leads 
to increasing probability for configurations with four images.
Furthermore, large ellipticities ($e> 0.606$) or shears
($\gamma > 1/3$) can lead to ``naked cusp'' configurations with 
three bright images \citep[e.g.,][]{KKS}.  Nearly all known lenses 
with point-like images are doubles or quadruples; among $\sim$80 
known lenses there is only one candidate naked cusp lens 
\citep[APM~08279+5255;][]{lewis02}.

\begin{figure}
\centerline{ \epsfig{file=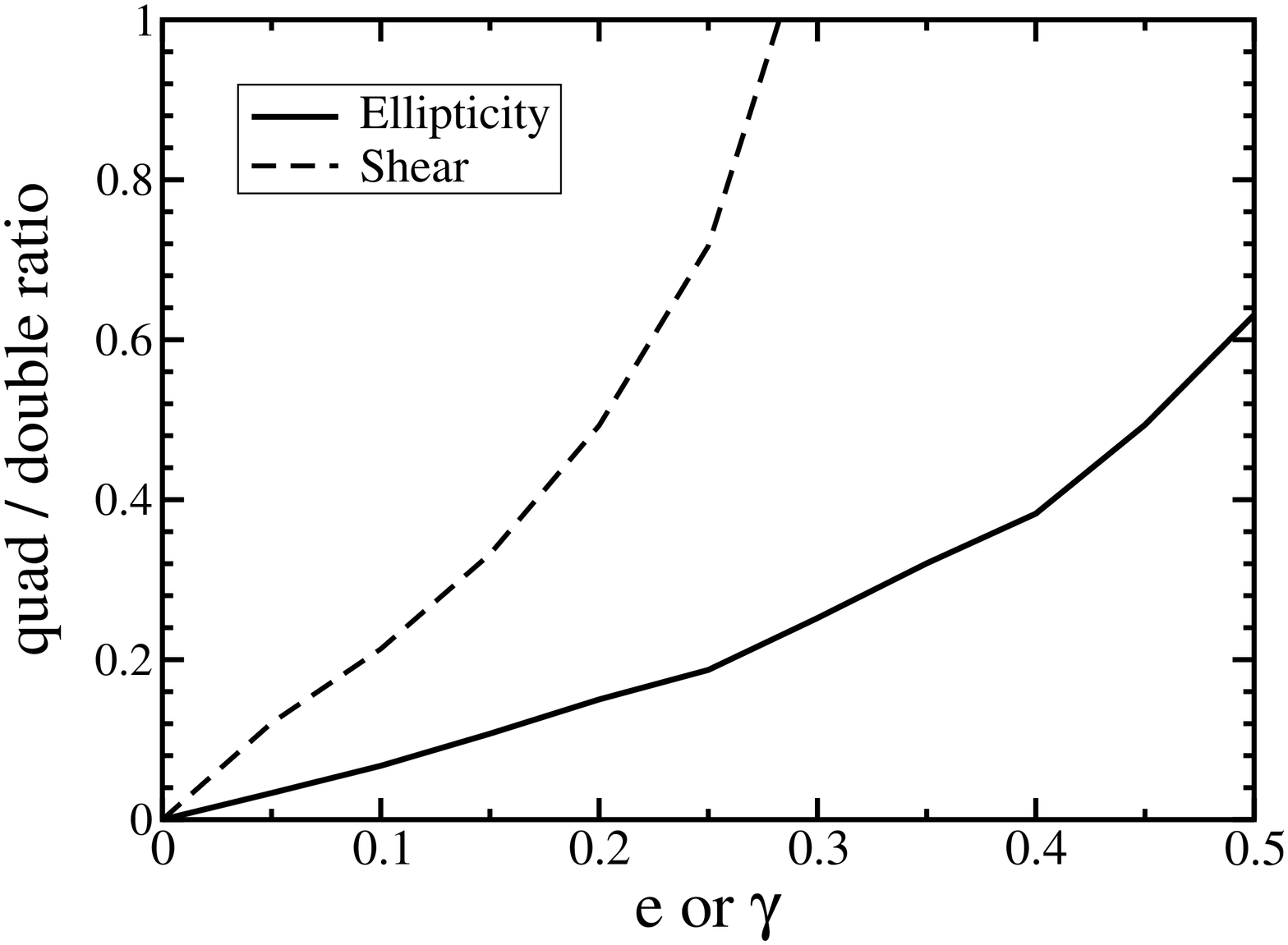, height=3.5in,
width=2.5in, angle=-90} }
\caption{
Quadruple-to-double ratio as a function of ellipticity or shear, 
assuming the CLASS LF (a power law with $\beta=2.1$).
}\label{fig:qd1}
\end{figure}

\reffig{qd1} shows that the quadruple to double ratio rises 
monotonically with ellipticity or shear.  For the CLASS LF, the 
ratio is $\sim$20\% for typical ellipticities $e \sim 0.3$ or 
shears $\gamma \sim 0.1$.  Our results agree well with previous 
analyses \citep[e.g.,][]{rusin2,finch}. 

\begin{figure*}
\centerline{
\epsfig{file=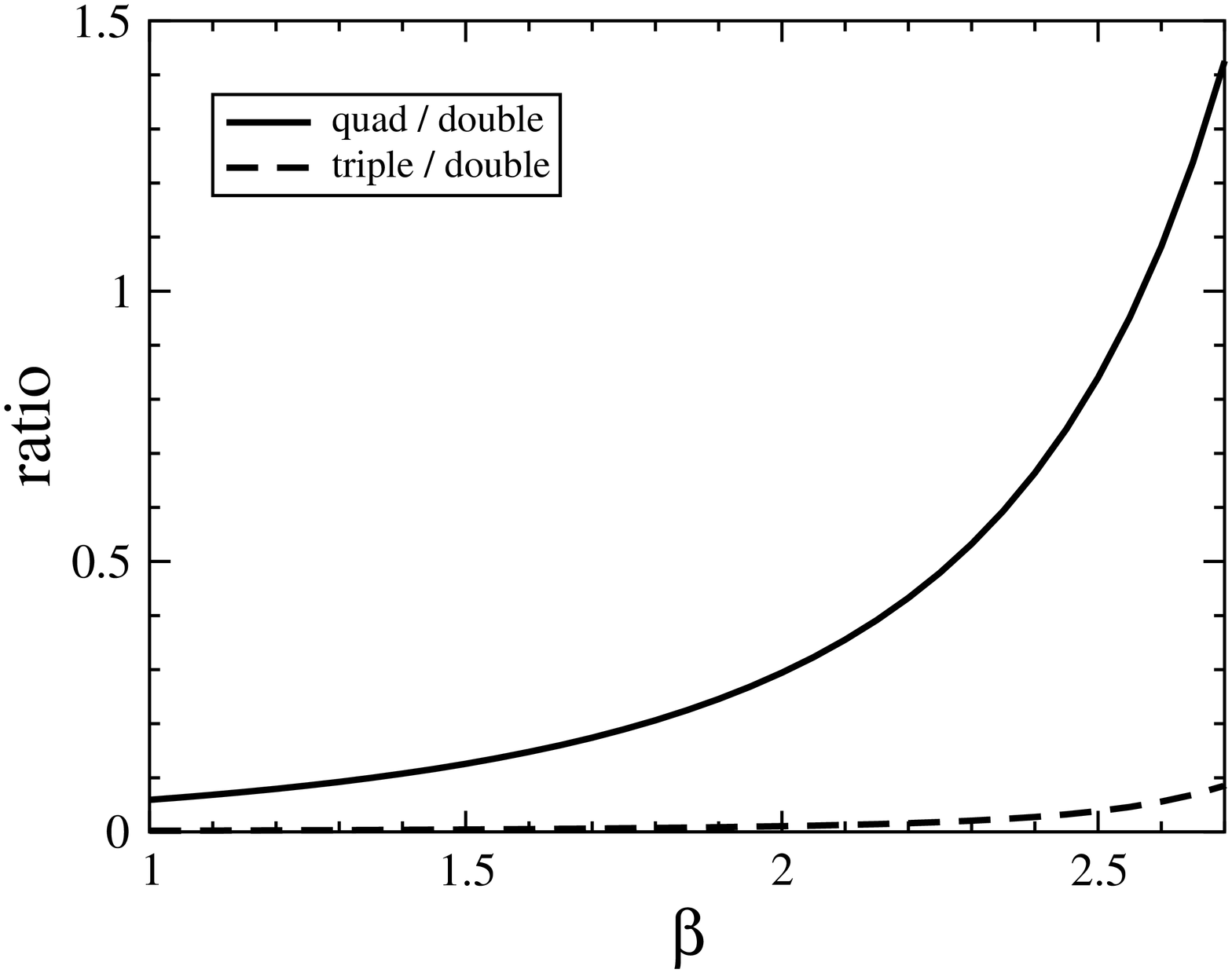, height=3.5in, width=2.5in, angle=-90}
\epsfig{file=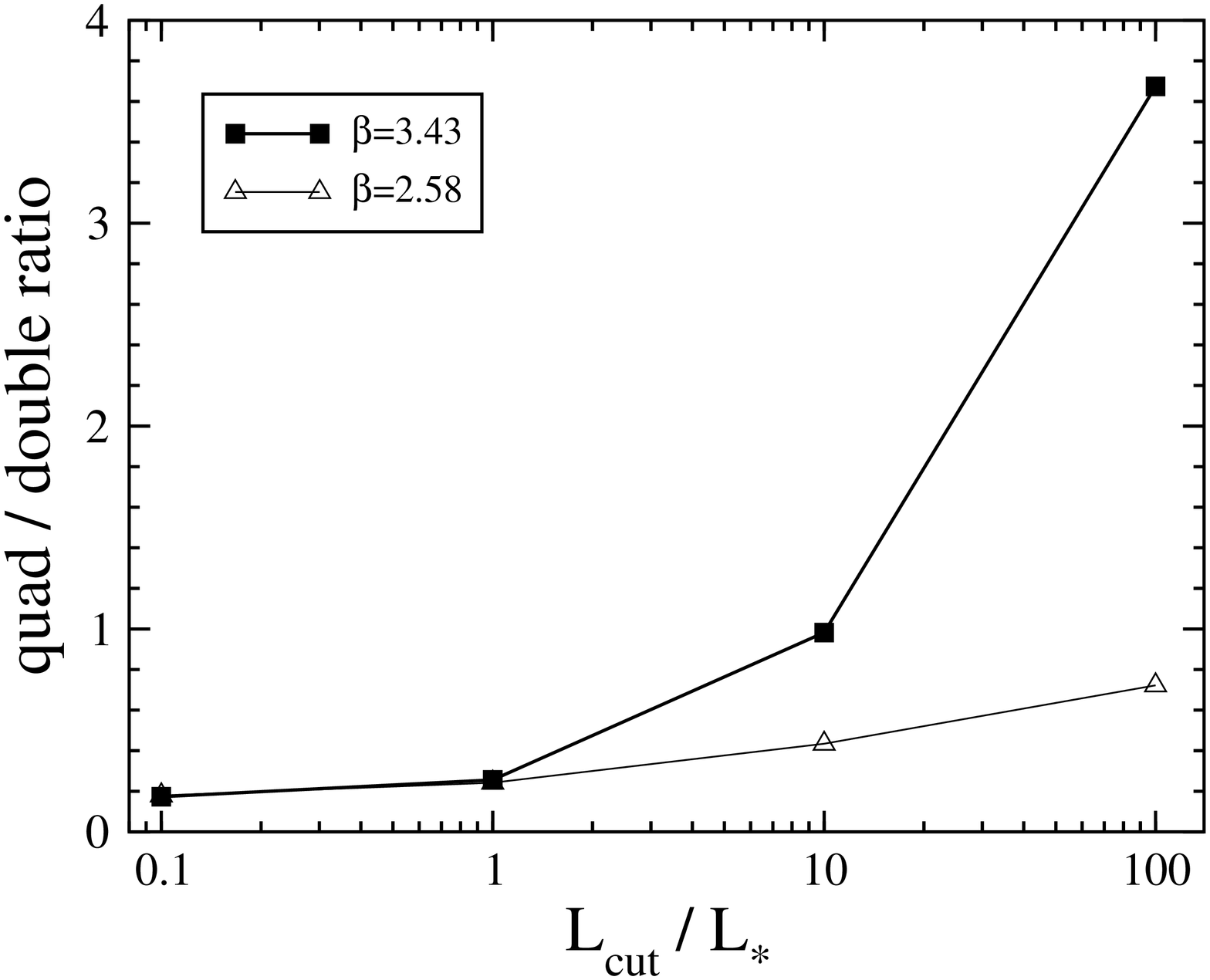, height=3.5in, width=2.5in, angle=-90}
}
\caption{ 
Quadruple-to-double ratio, as a function of the slope $\beta$ of 
power law LFs (left) or the limiting luminosity $\Lcut/L_*$ for 
quasar LFs (right).  The results are obtained by averaging over 
the distributions of ellipticity and shear discussed in the text.  
In the left panel, we also show the ratio of triple (or naked 
cusp) lenses as the dashed curve.
}\label{fig:qd2}
\end{figure*}

We can now estimate the expected number of quadruples (and cusp 
triples) by averaging over our fiducial distribution of 
ellipticity and shear.  The results are shown in \reffig{qd2} for 
both the power law LF and the quasar LF.  For the CLASS LF 
($\beta=2.1$), the net quadruple-to-double ratio is about 0.35, 
while the triple-to-double ratio is only about 0.01.  The number 
of quadruple vs.\ double systems in the CLASS statistical sample 
is 5 vs.\ 7; the ratio is twice as large as our prediction.  We 
therefore agree with \citet{rusin2} in concluding that ellipticity 
and shear alone cannot easily explain the high number of 
quadruples.  Additional effects are required, which are probably 
related to lens galaxy environments.  Shear is only a low-order 
approximation to the lensing effects of objects near the lens 
galaxy.  Recent studies have shown that including higher-order 
effects from satellite galaxies \citep{cohn2} or extended groups 
of galaxies \citep{KZ04} around the lens can significantly boost 
the quadruple-to-double ratio.

\reffig{qd2} shows that surveys targeting lensed quasars are 
expected to have a low quadruple-to-double ratio unless the bright 
end of the LF is steep ($\beta_h=3.43$) and the survey is limited 
to bright quasars ($\Lcut/L_* \gtrsim 10$).  This prediction 
could, of course, be an underestimate because we have neglected 
the higher order effects from lens environments.

\section{Effects on Cosmological Constraints}
\label{sec:cosmol}

While the changes in the optical depth and image separation 
distribution caused by ellipticity and shear seem mild, it is 
important to quantify how they affect one of the main applications 
of lens statistics: constraints on cosmological parameters.  One 
approach would be to modify the analyses of real lens samples to 
include the full effects of ellipticity and shear (building upon 
the analysis of \citealt{chae}).  Such an approach, however, would 
be limited by Poisson uncertainties in current lens samples (e.g., 
CLASS has just 13 lenses), by systematic uncertainties where 
models may or may not be correct (e.g., evolution in the lens 
galaxy population; see \citealt{mitchell}), and by systematic 
effects that are known to be present in the data but have not yet 
been studied (e.g., having multiple lens galaxies).  We believe 
that it is more instructive to create mock lens surveys that mimic 
CLASS but allow us to isolate the effects of ellipticity and 
shear.  Specifically, we create surveys that include ellipticity 
and shear, and then analyze them using standard spherical models 
in order to uncover biases that result from neglecting ellipticity 
and shear.  We create mock surveys with 1000 lenses in order to 
minimize Poisson uncertainties.  We use a Monte Carlo approach, 
drawing parameter values from appropriate probability 
distributions (as indicated in eq.~\ref{eq:tau}).  Specifically:
\begin{itemize}

\item
A subset of sources in the CLASS survey has a redshift  
distribution that can be treated as a Gaussian with mean 
$\avg{z_s}=1.27$ and width $\sigma_z=0.95$ \citep{marlow}, and 
this is usually taken as a model for the redshift distribution of 
the full survey \citep[e.g.,][]{chae}.  The Gaussian is modified 
by the redshift dependence of the optical depth $\tau(z_s)$ to 
obtain the redshift distribution of the lensed sources 
\citep[see][]{mitchell}.

\item
The lens redshift is drawn from the distribution
$p(z_l) \propto (D_{ls}/D_{os})^2\ dV/dz_l$, where the factor of 
$(D_{ls}/D_{os})^2$ comes from the factor of $\theta_E^2$ in the 
lensing cross section (see \refeq{SISb}).

\item
The velocity dispersion is drawn from
$p(\sigma) \propto \sigma^4\ dn/d\sigma$, where the factor of 
$\sigma^4$ comes from $\theta_E^2$ in the lensing cross section.  
We use the velocity dispersion distribution function $dn/d\sigma$ 
derived by \citet{sheth} for early-type galaxies in the Sloan 
Digital Sky Survey.  For simplicity, we assume that the velocity 
dispersion distribution does not evolve with redshift.  We could 
add evolution to both the creation and analysis of the mock survey 
\citep[see, e.g.,][]{mitchell}, but that would just complicate 
matters.

\item
We use the ellipticity distribution from \citet{jorgensen}, the 
shear amplitude distribution from \citet{holder}, and random shear 
directions.

\item
When drawing random source positions, we use magnification bias 
appropriate to the CLASS survey (a power law with $\beta=2.1$) 
since it is the most commonly used survey in current lens 
statistic analyses.

\end{itemize}
Given the parameters we can compute the observables for each mock 
lens: the source and lens redshifts and the image separation.  The 
other key observable is the total number of sources in the survey.  
We use the optical depth to determine the number of sources needed 
to obtain 1000 lenses, which is typically $\sim\!8\times10^{5}$.  
We distribute these sources in redshift using the Gaussian given 
above.

We then analyze the mock survey with standard maximum likelihood 
techniques.  Assuming complete data --- knowledge of the image
separation and the lens and source redshifts for lens systems,
and the redshift distribution of non-lensed sources --- we use
the likelihood function
\begin{equation}
  {\cal L} = \frac{(N_{\rm pred})^{N_{\rm obs}}\ 
    e^{-N_{\rm pred}}}
    {N_{\rm obs}!} \times
    \prod_{i=1}^{N_{\rm lens}} \frac{1}{\tau(z_{s,i})}\,
    \frac{\partial^2\tau}{\partial z_{l,i}\ \partial\dt_i}\ .
\end{equation}
The first term represents the Poisson probability for having 
$N_{\rm obs}$ observed lenses when $N_{\rm pred}$ are predicted, 
while the second term represents the probability that the lenses 
have the observed properties (e.g., observed lens redshift $z_l$ 
and image separation $\dt$ given the source redshift $z_s$).  As 
mentioned above, we neglect ellipticity and shear in the 
likelihood analysis because we want to understand the biases that 
may occur when spherical models are used for lens statistic 
analyses.  We hold the parameters in the velocity dispersion 
function fixed at their input values since uncertainties in these 
parameters have negligible effect \citep{mitchell}.  Thus, the 
only variables in the model are the cosmological parameters 
$\Omega_M$ and $\Omega_\Lambda$, which we adjust to maximize the 
likelihood.  We use input values of $\Omega_M = 0.3$ and 
$\Omega_\Lambda = 0.7$, and study how much the recovered values 
differ.  As mentioned above, using surveys with 1000 lenses should 
mitigate Poisson uncertainties, but we always produce and analyze 
10 independent surveys to verify that the statistical noise in our 
results is negligible.

\begin{figure}
\centerline{\epsfig{file=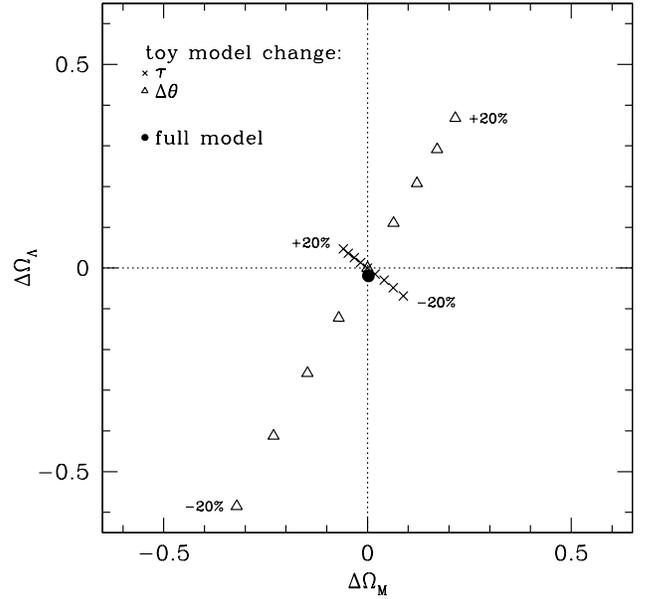, width=3.3in}}
\caption{ 
Biases in constraints on cosmological parameters from analyses of 
lens statistics.  We show the errors
$\Delta\Omega_M = \Omega_M^{\rm mod} - \Omega_M^{\rm true}$ and 
$\Delta\Omega_\Lambda = \Omega_\Lambda^{\rm mod} – 
\Omega_\Lambda^{\rm true}$ that result from using simple models 
with spherical lenses that neglect shifts in the optical depth 
 (crosses), image separations (triangles), or both.  (See text for 
details.)  The statistical uncertainties are smaller than the size 
of the points.
}\label{fig:cosmol}
\end{figure}

It is useful to begin by examining two toy models that focus on 
how changes in the optical depth or image separation distribution 
can affect cosmological constraints.  In the first case, we 
imagine using spherical lens models but manually adjusting the 
optical depth.  This is equivalent to changing the total number of 
deflectors.  In practice, it means adjusting the number of sources 
in our mock survey (since we fix the number of lenses).  The 
crosses in \reffig{cosmol} show the errors in the recovered 
cosmological parameters if the difference between the actual 
(input) optical depth and the spherical model is
$(-20,-15,-10,-5,0,5,10,15,20)\%$.  We see that simply changing 
the optical depth moves the cosmological parameters mainly along 
the line corresponding to flat cosmologies, and the shift is 
fairly small: $\Delta\Omega_\Lambda = 0.03$ if the real optical 
depth is 10\% larger than predicted by the spherical model.

In the second case, we again start with spherical models but 
manually adjust the image separations.  This is equivalent to 
shifting the velocity dispersion distribution to higher or lower 
values, and then adjusting the number of galaxies to keep the 
optical depth fixed.  The triangles in \reffig{cosmol} show the 
results of shifting the image separations by
$(-20,-15,-10,-5,0,5,10,15,20)\%$.  There is a large shift in the 
recovered cosmological parameters, and it is almost orthogonal to 
the line of flat cosmologies.  For example, if the real image 
separations are 10\% larger than predicted by spherical models, 
then there will be errors of $\Delta\Omega_M = 0.12$ and 
$\Delta\Omega_\Lambda = 0.21$ in the parameters recovered by 
spherical models.  These two cases are just toy examples, but they 
illustrate the important principle that even small errors in the 
model image separations can have a significant effect on 
cosmological constraints (even if small errors in the optical 
depth do not).

Finally, we consider the case where we use the full effects of 
ellipticity and shear on the mock survey.  Essentially, this 
amounts to using the corrections to the optical depth from 
\reffig{tauavg} and to the image separation distribution from 
\reffig{sep3}.  The circle in \reffig{cosmol} shows that 
neglecting ellipticity and shear in the likelihood analysis causes 
errors of $\Delta\Omega_M = 0.00\pm0.01$ and $\Delta\Omega_\Lambda 
= -0.02\pm0.01$, where the errorbars represent the statistical 
uncertainties in our calculations.  (We have achieved small 
Poisson uncertainties but not eliminated them altogether.)  That 
is the case if we allow $\Omega_M$ and $\Omega_\Lambda$ to vary 
independently.  If we restrict attention to flat cosmologies 
($\Omega_M + \Omega_{\Lambda} = 1$) then the bias is just 
$\Delta\Omega_M = -\Delta\Omega_\Lambda = 0.01$ (with negligible 
errorbars).  This result is consistent with our conclusions from 
the previous sections that ellipticity and shear have little 
effect on the optical depth and mean image separation.  It is 
nonetheless valuable to have a careful validation of the 
conventional wisdom that ellipticity and shear do not 
significantly affect cosmological constraints derived from lens 
statistics.

\section{Conclusions}
\label{sec:concl}

The effects of ellipticity and shear on strong lensing statistics 
have been swept under the rug in most analyses to date (a valiant 
exception being \citealt{chae}).  The reason for this is twofold: 
(1) models with nonspherical deflectors introduce new, and 
sometimes poorly constrained, parameters and greatly complicate 
calculations; and (2) conventional wisdom suggested that realistic 
ellipticities and shears had little effect on anything but the 
image multiplicities.  We have stepped beyond the state of 
blissful ignorance to present a general analysis of how 
ellipticity and shear enter into lens statistics.

The effects depend strongly on magnification bias, which in turn 
depends on the luminosity function of sources in a lens survey.   
If the LF is a power law $\propto L^{-\beta}$, as in radio surveys 
like CLASS (with $\beta \approx 2.1$), ellipticity and shear 
generally decrease or increase the lensing optical depth by only a 
few percent.  The increase is more than 5\% only if the LF is 
steep ($\beta \gtrsim 2.5$).  For optical quasar surveys, if the 
limiting luminosity is below the break in the quasar LF then 
ellipticity and shear decrease the optical depth by a few percent.  
There is a noticeable ($>$5\%) increase only for surveys limited 
to the brightest quasars ($\Lcut/L_* \gtrsim 10$, if the bright 
end slope is $\beta_h = 3.43$).  Since ongoing and planned optical 
surveys are expected to reach to the faint end of the QSO LF 
($\Lcut/L_* \lesssim 1$), we do not expect ellipticity and shear 
to have a large effect on the predicted number of lenses in future 
lens surveys.

Ellipticity and shear do not shift the mean of the distribution of 
lens image separations, but they do introduce an additional 
scatter of $\sim$10\%.  They naturally affect the relative numbers 
of double, quadruple, and triple lenses, but they cannot easily 
explain the high observed quadruple-to-double ratio.
Ellipticity has little effect on predictions for elusive 
central lensed images, although it does lead to a segregation that 
quadruple lenses are generally expected to have fainter central 
images than double lenses \citep{CoreImg}.

Since ellipticity and shear produce only small changes in the lensing
optical depth and image separation distribution, they are not very
important in lensing constraints on cosmological parameters.
Neglecting them leads to biases in $\Omega_M$ and $\Omega_\Lambda$ of
$<$0.02.  Moreover, hydrodynamical N-body simulations tend to find
systems that are more spherical than those in dissipationless
simulations \citep[e.g.,][]{stelios}. Therefore, the ellipticity
effects on lensing statistics found in this paper, while already
small, could even be an overestimate.

We conclude that for lens statistics problems other than image 
multiplicities, ellipticity and shear have surprisingly little 
effect.  Unless percent-level precision is needed, or a survey 
with a particularly steep LF is being considered, ellipticity and 
shear can probably be ignored.   Their effects will become 
more important as lens samples grow into the hundreds or thousands 
and statistical uncertainties plummet \citep[see, e.g.,][]{KKM}.  
At that time it will be important to know the distributions of 
ellipticity and shear, and also to resolve questions about how to 
normalize the lens models (see \refsec{SIEg}).

There are systematics besides ellipticity and shear that may affect
strong lens statistics.  They include mergers and evolution in the
deflector population
\citep[e.g.,][]{rix94,mao-koch,keeton02,ofek,chae-evol,mitchell},
halo triaxiality \citep[e.g.,][]{oguri2} or other complex internal
structure \citep[e.g.,][]{moller3,Qua03}, compound lens galaxies
\citep[e.g.,][]{csk88,moller,cohn2}, and lens galaxy environments
\citep[e.g.,][]{moller2,KZ04}.  In order to bring lens statistics
into the realm of precision cosmology, each of these factors must
be addressed carefully.  We have taken one step in that direction
by studying ellipticity and shear, finding that their effects are
relatively small and in principle easy to take into account.

\acknowledgements

We thank Andrey Kravtsov for interesting discussions that prompted 
us to examine the biases in cosmological parameters.  We thank 
Chris Kochanek and the anonymous referee for good questions about 
the model normalization.
DH is supported by the DOE grant to CWRU. 
CRK is supported by NASA through Hubble Fellowship grant
HST-HF-01141.01-A from the Space Telescope Science Institute,
which is operated by the Association of Universities for Research
in Astronomy, Inc., under NASA contract NAS5-26555.
C-P Ma is supported by NASA grant NAG5-12173 and a Cottrell 
Scholars Award from the Research Corporation.

\appendix

\section{Why is $A/A_0 \le 1$?}

In this appendix we derive the cross section for a generalized 
isothermal lens to explain the result from \refsec{tau} that 
ellipticity reduces the cross section.  The lens potential for a 
generalized isothermal model has the form
$\Phi_{\rm iso}(r,\phi) = r\,f(\phi)$ where $f(\phi)$ is an 
arbitrary function specifying the angular shape.  Consider 
expanding the potential in angular multipoles,
\begin{equation}
  \Phi_{\rm iso}(r,\phi) = \theta_E\,r\,\left( 1 -
    \sum_{m=1}^{\infty} \Bigl[ a_m \cos(m\phi) + b_m \sin(m\phi) 
\Bigr]
    \right) ,
\end{equation}
where $\theta_E$ is the Einstein radius (as defined in  
\refsec{SIEg}).
The corresponding mass distribution then has the form
\begin{equation}
  \kappa_{\rm iso}(r,\phi) = \frac{\theta_E}{2r} \left( 1 +
    \sum_{m=1}^{\infty} \epsilon_m \cos[m(\phi-\phi_m)] \right) ,
\end{equation}
where the amplitude $\epsilon_m$ and direction $\phi_m$ of the 
mass multipole are given by
\begin{eqnarray}
  \epsilon_m &=& (m^2-1) \sqrt{a_m^2+b_m^2}\ , \\
  \phi_m &=& \frac{1}{m} \tan^{-1}\frac{b_m}{a_m}\ .
\end{eqnarray}
In other words, we can think of this model as a multipole 
expansion of the surface mass density.

\begin{figure}
\centerline{\epsfig{file=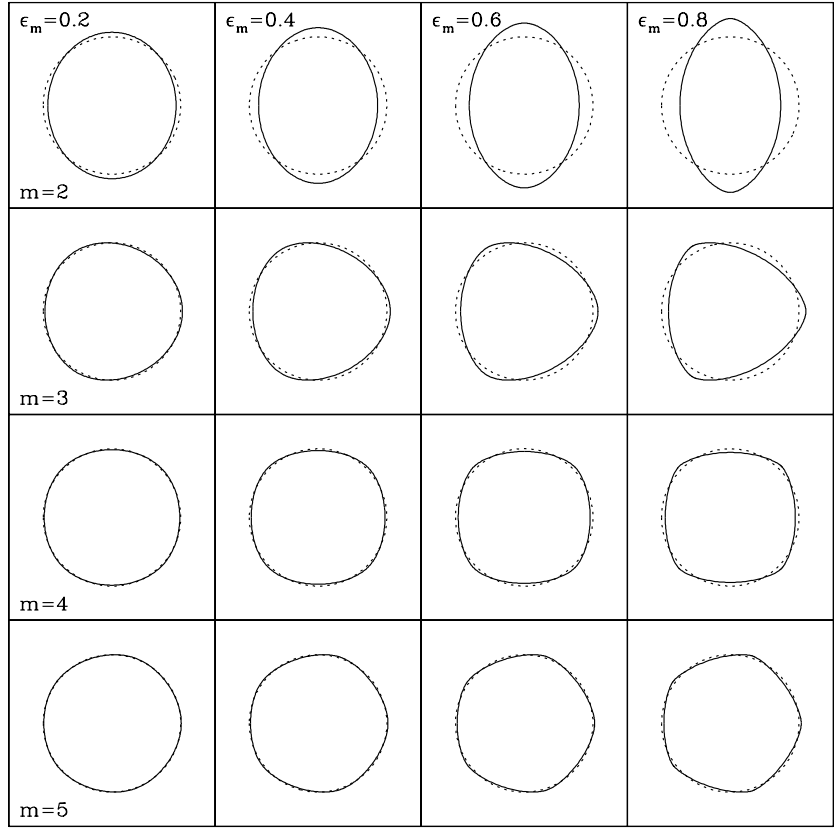, height=4.5in}}
\caption{
Radial caustics for isothermal galaxies with different multipole 
moments.  In each panel, the solid curve shows the caustic for the 
specified model, while the dotted curve shows the caustic of a 
spherical model for reference.  The order $m$ of the multipole 
moment increases from top to bottom, and the amplitude 
$\epsilon_m$ increases from left to right.
}\label{fig:caus}
\end{figure}

The radial caustic --- properly termed a pseudo-caustic since a 
singular isothermal lens does not formally have a radial critical 
curve \citep[see][]{evans} --- can then be written in parametric 
form as:
\begin{eqnarray}
  u_{\rm caus}(\lambda) &=& -\theta_E \left( \cos\lambda +
    \sum_{m=1}^{\infty} \Bigl[
        (a_m \cos m\lambda + b_m \sin m\lambda)\cos\lambda +
      m (a_m \sin m\lambda - b_m \cos m\lambda)\sin\lambda
    \Bigr] \right) \\
  v_{\rm caus}(\lambda) &=& -\theta_E \left( \sin\lambda +
    \sum_{m=1}^{\infty} \Bigl[
        (a_m \cos m\lambda + b_m \sin m\lambda)\sin\lambda -
      m (a_m \sin m\lambda - b_m \cos m\lambda)\cos\lambda
    \Bigr] \right)
\end{eqnarray}
Although this form appears complicated, if we collect the two 
coordinates $u_{\rm caus}$ and $v_{\rm caus}$ into a vector 
$\u_{\rm caus}$ then we can easily evaluate the area inside the 
radial caustic:
\begin{equation}
  A = \int_{0}^{2\pi} \frac{1}{2} \left[ \u_{\rm caus}(\lambda) 
    \times \frac{d\u_{\rm caus}}{d\lambda} \right] d\lambda
  = \pi\,\theta_E^2 \left[ 1 - \frac{1}{2} \sum_{m=1}^{\infty}
    (a_m^2+b_m^2) (m^2-1) \right] .
\end{equation}
This is the lensing cross section (provided there are no naked 
cusps).  The summand, and hence the sum, is manifestly 
nonnegative, so the cross section for {\it any\/} nonspherical 
model is $A < A_0 \equiv \pi \theta_E^2$.  This result is 
illustrated in \reffig{caus} for different multipole terms.  It is 
clear that asphericity deforms the caustics in such a way that the 
cross section is smaller than for the spherical case.

\end{document}